\documentclass[fleqn,10pt]{wlscirep}
\usepackage[utf8]{inputenc}
\usepackage[T1]{fontenc}

\title{Far-ranging generalist top predators enhance the stability of meta-foodwebs}

\author[1,*]{Andreas Brechtel}
\author[2]{Thilo Gross}
\author[1]{Barbara Drossel}

\affil[1]{Technische Universität Darmstadt, Institute for condensed matter physics, Darmstadt, Germany}
\affil[2]{UC Davis, Department of Computer Science, 1 Shields Av, Davis, Ca 95616, USA}
\affil[*]{brechtel@fkp.tu-darmstadt.de}
\begin{abstract}
Identifying stabilizing factors in foodwebs is a long standing challenge with wide implications for community ecology and conservation. 
Here, we investigate the stability of spatially resolved meta-foodwebs with far-ranging super-predators for whom the whole meta-foodwebs appears to be a single habitat. 
By using a combination of generalised modeling with a master stability function approach, we are able to  efficiently explore the asymptotic stability of large classes of realistic many-patch meta-foodwebs.
We show that meta-foodwebs with far-ranging top predators are more stable than those with localized top predators. Moreover, adding far-ranging generalist top predators to a system can have a net stabilizing effect, despite increasing the food web size. These results highlight the importance of top predator conservation. 
\end{abstract}

\begin{document}

\flushbottom
\maketitle
\thispagestyle{empty}

\setlength{\parindent}{0ex}
\setlength{\parskip}{0.5em}

\section*{Introduction}
A persistent theme in community ecology is the quest to understand the factors that stabilize foodwebs. Between early notions of ``Complexity begets stability'', May's mathematical argument that more complex foodwebs should be less stable\cite{May1972}, and a number of recent explorations using simulations and stability analysis, a plethora of different hypotheses on foodweb stability have been explored. Among the most prominent of these is the stabilizing effect of weak links \cite{McCann1998,McCann2000}, although other evidence suggests a potentially destabilizing effect of these links \cite{Jansen2003, Gross2009}. Further stabilizing effects are ascribed to allometric scaling \cite{Kartascheff2010,Brose2006}, nontrophic interactions \cite{Kefi2015}, and certain foodweb motifs \cite{Neutel2002,Bascompte2005,Allesina2008,Gross2009}.       
The topic of foodweb stability has gained renewed urgency by recent papers raising concerns that past perturbations have not only led to the loss of species but left ecological systems in a significantly more fragile state  \cite{BarOz2009,Yeakel2014}. A likely candidate contributing to this fragility of previously disturbed systems is the loss of top predators. 
Habitat loss, overexploitation, and numerous other stress factors have caused global declines in top predators \cite{Stiere2016}. For example in  the British isle all of the large top predators such as the wolf, bear, and lynx have gone extinct.

Among the species in a foodweb top predators appear to have a disproportionate impact on stability. Althoug they constitute typically only a tiny fraction of the biomass \cite{Brose2017}, top-predators have been found to have a strong impact on stability \cite{Parsons1992,Gross2009,Ripple2014}. Top predators are often generalists which connect and balance biomass from different types of specialist predators on lower trophic levels \cite{Post2000}. Thereby they also act at the apex point of weakly-linked long loops in foodwebs which are thought to be a stabilizing motif \cite{Neutel2002}.     

A less studied but perhaps equally important factor contributing to the role of top predators is the spatial nature of ecological systems. The effect of space has been studied in ecology for a long time in the context of classical metapopulation models \cite{hanski1999metapopulation,Taylor1990}, island biogeography \cite{macarthur2001theory}, and movement ecology \cite{Nathan2008}. However, spatially explicit studies of the dynamics of large foodwebs (meta-foodwebs) have only begun to appear recently \cite{Pillai2010,Ristl2014,Gravel2016,Mougi2016,Barter2016,Pilosof2017,Hamm2017,Brechtel2018}. Based on earlier studies \cite{Brown1977,Levins1969,muneepeerakul2008neutral,Levin} and the recent work of Pillai et al. \cite{Pillai2010,Pillai2011} we can expect that complex spatial environments aid the persistence of diverse communities by increasing the opportunities for generalists and enabling quick recovery from localized perturbations via the rescue effect.

In addition to beneficial consequences, spatial environments can also create intrinsic instabilities leading to complex and potentially dangerous spatiotemporal dynamics \cite{Baurmann2007, Gramlich2015, Brechtel2018}. 
In \cite{Brechtel2018} we introduced an efficient method for identifying such instabilities in large spatially explicit models of foodwebs dynamics, by combining a generalized modelling approach \cite{Gross2006,Yeakel2011} with a Master-Stability function method \cite{Pecora1998,Boccaletti2014}. 

A feature that is not captured in previous models is that top predators perceive the habitat differently than other species. Top predators tend to be large-bodied highly mobile species that need to roam large areas to cover their energy needs. While a landscape may appear as a complex network of many habitat patches from the perspective of a mouse it only consists of a single patch from the perspective of an eagle circling overhead. 

The top-predator's different perspective on the landscape is interesting because it means that the top predator cannot only link different biomass streams and close long loops in foodwebs by feeding on different species, it can also do so by simultaneously preying on populations in different habitat patches. 
  
In the present paper we investigate the stability of meta-foodwebs that appear as a complex network of habitat patches to all but the top predator population, while  
the top predator regards the whole system as a single \emph{global} patch.
We refer to such ultra-mobile predators as global predators from now on. To study this system we use a generalized modelling approach and extend the master-stability function approach to meta-foodwebs \cite{Brechtel2018} to accommodate a global top-predator. 

We find that systems with a global top predator tend to be more stable than systems with the same number of species where all species are local. 
This effect is particularly pronounced when the global predator feeds on several species, and thus acts as a generalist. In this case the stability of the system with a global top predator exceeds the stability of the smaller system where the global predator is absent. 
Our results highlight the importance of global generalist top predators in safeguarding the systems stability against disturbances.

\section*{Model}
We consider a system of $S$ local species on $N$ patches that disperse between these patches according to a network of dispersal routes, and $S'$ global species that can feed on all patches simultaneously. Below we particularly consider $S'=1$ in examples, while in the mathematical work we leave $S'$ undefined for generality. We denote the biomass density of local species $i$ in patch $k$ by $X_i^k$ and the biomass density of the global species $i$ by $Y_i$. The dynamical equations of the model have the following form:
\begin{align}
\label{eqn:biomass_ode1}
\dot X_i^k
	&= G_{X_i^k}(X_i^k) - M_{X_i^k}(X_i^k) + \varepsilon_{X_i} F_{X_i^k}(\mathbf X^k) - \sum_j D_{X_j^k X_i^k}(\mathbf X^k) + \sum_l \left[ E_i^{kl}(\mathbf X^k, \mathbf X^l) - E_i^{lk}(\mathbf X^l, \mathbf X^k) \right] - \sum_j D_{Y_j X_i^k}(\mathbf X, \mathbf Y)\,, \\
    \label{eqn:biomass_ode2}
\dot Y_i &= G_{Y_i}(Y_i) - M_{Y_i}(Y_i) + \varepsilon_{Y_i} \sum_{i,k} F_{Y_i}(\mathbf X, \mathbf Y) - \sum_j D_{Y_j Y_i}(\mathbf X, \mathbf Y) \, ,
\end{align}
where $G$ is the rate of growth by primary production, $M$ the rate of the loss by respiration and mortality, $F$ the rate of growth due to feeding on other species, $D$ the rate of biomass loss due to being eaten by other species, and $E_i^{kl}$ the rate of dispersal from patch $l$ to $k$. Furthermore, $\epsilon$ denotes the conversion efficiency of prey biomass into predator biomass. The subscripts indicate the affected populations. If two populations are affected, the order of the subscripts is such that biomass flows from the second to the first population. For instance, the function $D_{Y_j X_i^k}$ describes the biomass loss of the local population of species $i$ on patch $k$  due to predation by the global species $j$.
For this generalized equation system, and without restricting the functions in the equation to a specific functional form, one can then derive a Jacobian matrix that describes the response of the system to perturbations. Because of the uncertainties that exist, this Jacobian matrix still contains a number of unknown (but directly interpretrable) parameters that describe the topology of the network, the scale of biomass turnover (scale parameters), and the strength of nonlinearities (exponent parameters).  

We use the niche model \cite{Niche} to generate foodwebs with realistic topologies. An example of a five-species niche web is shown in figure \ref{fig:local5}. 
We identify the niche value with the logarithm of the body mass, which allows us to implement allometric scaling of the biomass turnover rates. We then generate the scale and exponent parameters, taking the food web structure into account. The parameters are drawn at random from suitably defined intervals in accordance with ecological reasoning (see Methods, Tab.~\ref{tab:ModelParameters}) and Fig.~8. The resulting sets of the generalized parameters closely follow previous publications, such as \cite{Gross2006,Yeakel2011,Plitzko2012,Gramlich2015}.  

We use random geometric graphs (RGG)\cite{Gilbert1961} as a model for the spatial topology of the network of habitat patches (see Methods). Unless mentioned otherwise, we consider homogeneous systems, where all patches are characterized by the same parameters and have the same biomass densities. This assumption enables the construction of master stability functions (explained below) that allow us to efficiently study systems with arbitrary number of geographical habitat patches.

To construct the master stability function we first note that the Jacobian has the form
\begin{align}\label{eqn:general_jacobian}
	\mathbf J =
	\begin{pmatrix}
		\mathbf J_X & \mathbf J_{XY} \\
		\mathbf J_{YX} & \mathbf J_Y
	\end{pmatrix} \,,
\end{align}
where the indices $X$ and $Y$ refer again to local and global species. 
Furthermore, the block $\mathbf J_X$ has the internal structure
\begin{align}
	\mathbf J_X = \mathbf I \otimes \mathbf P_X - \mathbf L \otimes \mathbf C_X 
    \label{eqn:Jx_homogeneous_main_simple}
\end{align}
familiar from systems without global predator \cite{Brechtel2018}, provided that the feeding rates of the global predator depend linearly on prey biomass. 
Here, $\otimes$ is a Kronecker product, $\mathbf I$ is an identity matrix of suitable size, $\mathbf P_X $ is the Jacobian of an isolated patch, $\mathbf L$ is the Laplacian matrix describing diffusion on the patch network, and $\mathbf C_X $  is the linearization of the dispersal terms \cite{Brechtel2018}. More details, and the generalization to global predators with nonlinear feeding rates, are given in the Methods section. 
  By exploiting the structure of the Jacobian its eigenvalues can be calculated by first finding the eigenvalues of the Laplacian matrix of the spatial network, and then computing the eigenvalues of a greatly reduced matrix in which the eigenvalues of the Laplacian appear as a parameter. 

While we confine the details of the eigenvalue construction to the methods section, we remark that the procedure follows the same idea as the master stability function approach used in the study of syncronization \cite{Pecora1998,Boccaletti2014} but is applied to steady states as in  
 \cite{Brechtel2018} and is further modified in this paper to accomodate the global predators.

The master stability function approach provides several advantages. First and foremost, it separates the contributions of the spatial structure to the dynamics from those of the local food web dynamics. This allows clearer ecological insight into the source of observed instabilities.  

Building on the separation of local and geographic structure one formulate precise conditions that the spatial structure must meet for a given food web to be stable. The function that represents these conditions is the actual master stability function after which this approach is named. 

A byproduct that we use here, is a drastic reduction in the numerical effort for finding the eigenvalues of the Jacobian. In the present paper this reduction enables us to obtain solid statistics in the numerical investigations of large many-patch many-species meta-foodwebs. 

To verify that the results described below also hold for systems with patches that violate the homogeneity assumption, we ran additional numerical analyses of systems with heterogeneous patches. In this case the spectrum of the Jacobian was determined by direct numerical diagonalisation of the full meta-foodweb Jacobian matrix. However, due to the large size of these matrices, we used smaller systems than for the homogeneous case. 

\section*{Results}
In the following we compare the mean stability of systems with a global top predator to that of systems with a local top predator or none at all.  
Then we evaluate how this stability depends on the sensitivity of the top predator to prey biomass, and on the number of prey populations.

\subsection*{Proportion of stable webs}
We quantify stability as the proportion of stable webs in an ensemble of systems that correspond to the same model. A web is stable if all eigenvalues of the Jacobian have a negative real part. 
An ensemble is characterized by the number of local species $S$, of global species $S'$ (which is 1 or 0), of spatial patches $N$, and a set of intervals for the generalized parameters, see Table 2. Each system of the ensemble is generated by randomly creating a spatial network (RGG), a niche foodweb with $S+S'$ species,
and a set of generalized parameters from the given intervals. For the foodwebs with a global predator the species with the highest niche value is selected as the global predator.

\begin{figure*}
	\centering
    \includegraphics{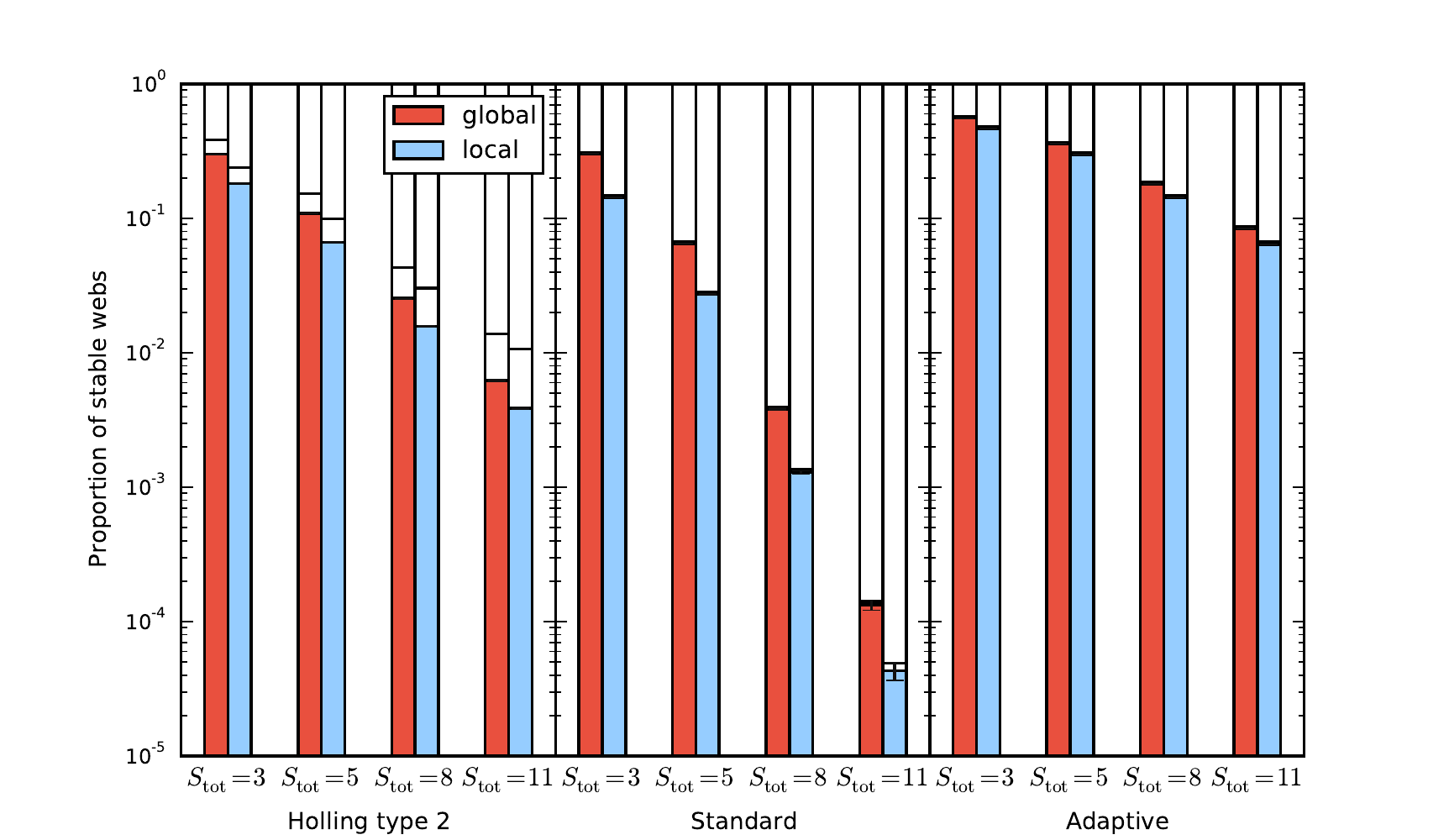}
	\caption{Proportion of stable webs for 10-patch systems with a global top predator (red) and systems with a local top predator (blue), for the three models specified by the parameter intervals listed in Table \ref{tab:ModelParameters}, and for different species numbers $S_{\rm{tot}}=S+S'$. The horizontal line separates the small proportion of unstable webs with the leading eigenvalue corresponding to a homogeneous mode from the larger proportion corresponding to an inhomogeneous (pattern-forming) mode. The figure shows that making a global predator local decreases stability and that the majority of instabilities is due to pattern-forming instabilities.}
	\label{fig:PSWbarplotHom}
\end{figure*}

\begin{figure*}
	\centering
    \includegraphics{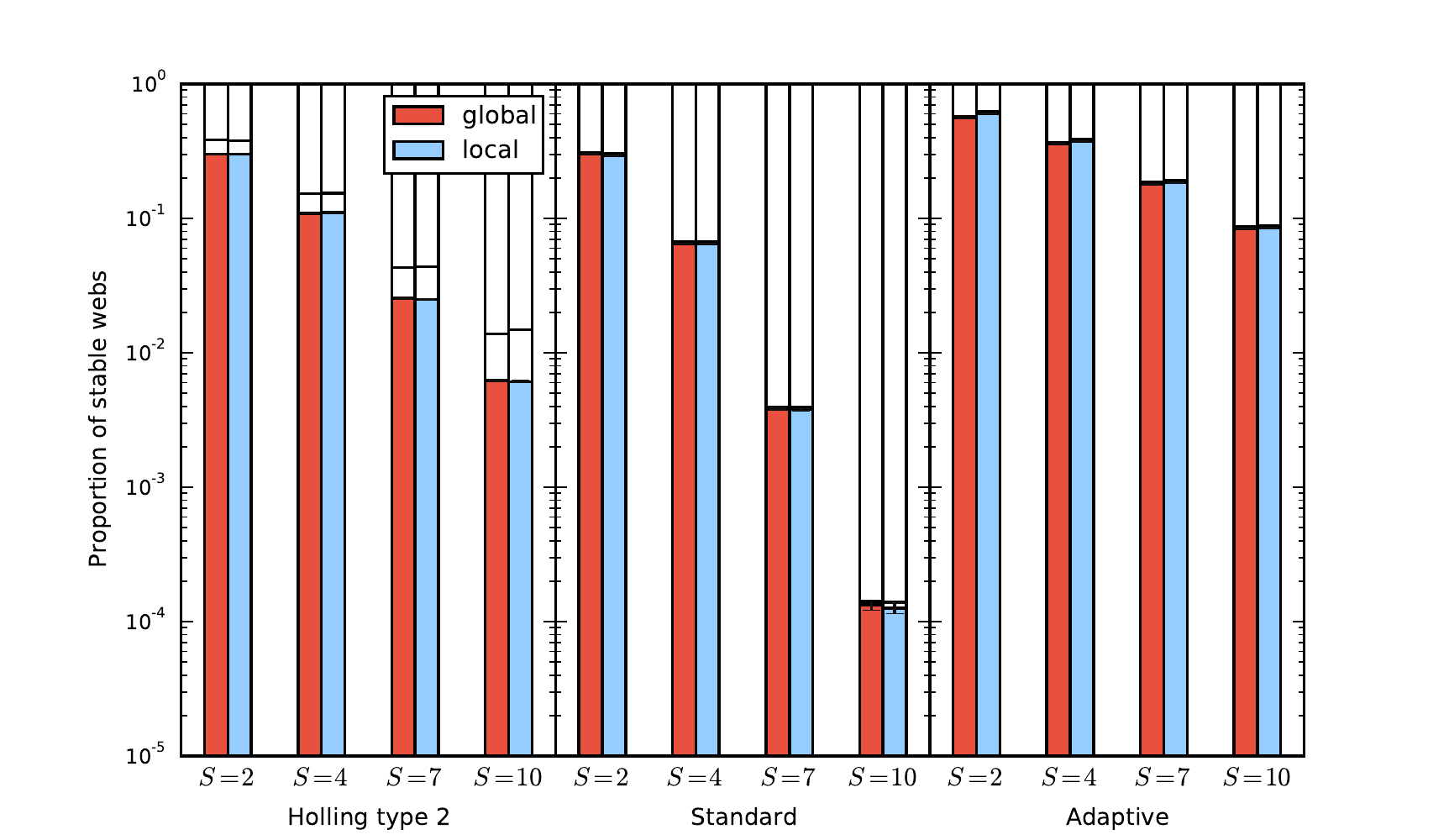}
	\caption{Same as figure~\ref{fig:PSWbarplotHom}, but in the system without the global predator no local predator is added. The proportions of stable webs and both types of instabilities are very similar in the systems with and without the global predator. Comparison with the previous figure shows that adding a global predator does not decrease stability while adding a local predator does decrease stability.}
	\label{fig:PSWbarplotHom2}
\end{figure*}

Figure~\ref{fig:PSWbarplotHom} shows a comparison of the proportion of stable webs of systems with the top predator added as a global species vs as a local species. The proportion of stable webs is in all examined cases larger for the systems with the global top predator than in the systems with the local one, with the increase ranging between 20\% and 200\%. For the unstable systems, we evaluated whether the leading unstable mode is homogeneous or inhomogeneous. Homogeneous instabilities make a minor contribution to the number unstable webs. This is because there are many more inhomogeneous than homogeneous eigenmodes. 
Figure~\ref{fig:PSWbarplotHom2} shows that the proportion of stable webs remains almost unchanged when the global top predator is removed from the system. 

Among the unstable systems, the proportion of homogeneous and inhomogeneous leading modes is also nearly the same. This is because the underlying eigenvalue problem for the stability of homogeneous systems can be separated into two types of solutions. One type represents the inhomogeneous eigenmodes where the global predator has the value 0 in the corresponding eigenvector. The other part represents the coupling of the global species to the spatial part of the system, which corresponds to the spatially homogeneous eigenvectors. When the global predator is removed, the spatially inhomogenoues part of the eigenvalue problem, which gives most of the eigenmodes, is essentially unchanged. Therefore the removal of the global predator has no strong impact on the proportion of stable webs. For more details on the eigenvalue problem see the Methods section.

To explore in more detail the influence of the global predator on stability we evaluate the change of stability with a key parameter related to predation by the top species, for systems with and without a global predator. This parameter, $\gamma$, describes the sensitivity of the top predator's food intake to the total available prey biomass. It is known that it has a positive correlation with stability\cite{Gross2006}. The reason is that the predator is close to saturation for small $\gamma$  and therefore can hardly control the population of its prey species.

\begin{figure*}
	\centering
	\includegraphics{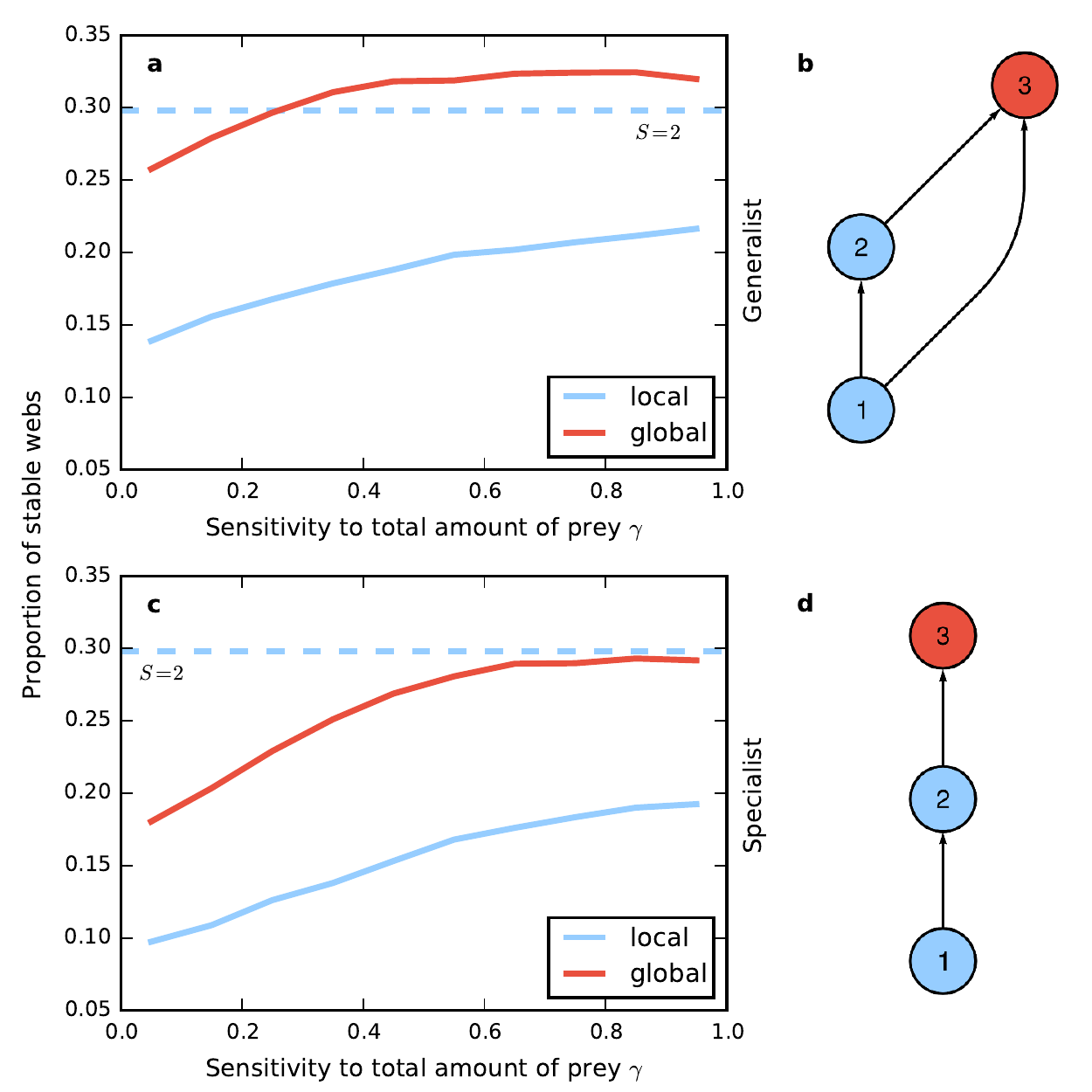}
	\caption{Proportion of stable webs for 10-patch Holling type 2 systems as function of the sensitivity to the total amount of prey $\gamma$ available to the top species (which is species number 3) of a 3-species omnivore module (top) and a 3-species food chain (bottom), with the top species being either global (red) or local (blue). Additionally, the mean proportion of stable webs of the 2 species system without the species 3 is shown as the horizontal dashed line. The data show that systems with global top species are more stable than those with local top species, and that the generalist module is more stable than the three-species food chain. With a generalist top species it is even more stable than the two-species system. }
	\label{fig:stab-gen-vs-spec}
\end{figure*}

At first, we use two types of three-species foodwebs, one with a generalist top species and one with a specialist top species (see figure~\ref{fig:stab-gen-vs-spec} b and d). We compare the situation that the top species is global with the situations where it is local or completely absent. In the latter case, we have a two-species system. The proportion of stable webs in dependence of the exponent parameter $\gamma$ of the top species for the two types of three-species systems is shown in figure~\ref{fig:stab-gen-vs-spec}. 

As before, the stability of the systems with the global top predator is higher than that with the local top predator. Furthermore, the stability of the systems with the generalist global top predator is also higher than that of the two-species system when $\gamma$ is larger than 0.25.  With the specialist top predator stability is smaller than for the two-species system but approaches it for larger $\gamma$. 

In the case of the three-species systems the generalist global species has an stabilizing effect, in contrast to the specialist one. However, from studying this small system it is not clear whether the stabilizing effect is due to the larger number of prey species or to the fact that the global species is an omnivore that feeds on different trophic levels. In order to investigate this issue, we also studied systems with more species, and we explored how stability depends on the number and trophic range of the top predator's prey species in these larger foodwebs.  Figure~\ref{fig:LinkNumberHist+NicheIntervalHist2D}~a shows the proportion of stable webs for system with $S=10$ local species and a global predator in dependence of the number of prey species of the global predator. A larger number of prey species is correlated with a higher proportion of stable webs. When the global predator has four prey species the stability reaches that of the system without global predator and exceeds it for larger numbers of prey species. 

In figure~\ref{fig:LinkNumberHist+NicheIntervalHist2D}~b the proportion of stable webs is shown in dependence of the niche interval size and the number of prey species of the global predator. The niche interval size is defined as the difference between the highest and the smallest niche values of the global predator's prey. Larger niche interval size implies that the global predator feeds on more trophic levels on average.  The figure shows that the global predator has the strongest stabilizing effect when it feeds on a large number of species with relatively similar niche values. This means that feeding on different trophic levels is not a stabilizing factor in our system. 

\subsection*{Heterogeneous systems}
The results presented so far were obtained using a spatially homogeneous model. Making the systems heterogeneous makes them more realistic as natural systems always have heterogeneities. Since we cannot use the Master Stability Function approach any more, the numerical effort becomes much larger, as we have to perform a direct numerical evaluation of the eigenvalues.   Heterogeneity is introduced by assigning the exponent parameters independently for every patch from the same uniform distributions as before (table~\ref{tab:ModelParameters}). We use the same foodweb structure in every patch. Hence we get the same scale parameters in every patch. Note that our way of constructing heterogeneous systems ignores correlations between the exponent parameters in different patches and is therefore a stronger type of heterogeneity than we would expect from nature. 

Figure~\ref{fig:PSWbarplotHet} shows the  proportion of stable webs. It is analogous to figure \ref{fig:PSWbarplotHom}, but for smaller, heterogeneous five-patch systems. For comparison the figure also includes the data for homogeneous webs of the same size as the heterogeneous ones. 
One obvious difference between the heterogeneous and homogeneous systems is the overall lower proportion of stable webs, but this need not always be the case. Previous work has shown that heterogeneity can increase as well as decrease linear stability in generalized models \cite{Gramlich2018}.

Figure~\ref{fig:LinkNumberHistHet} shows the dependence of stability on the number of prey of the global predator. These results are analogous to those of  figure~\ref{fig:LinkNumberHist+NicheIntervalHist2D}, but for a heterogeneous system with $S=7$ local species. The plot shows the same trend as for the homogeneous system. A larger number of prey species of the global predator leads to a higher proportion of stable webs. Furthermore, the proportion of stable webs is always above that obtained without global predator. In our example system, the stabilizing effect of the global predator is thus stronger in the heterogeneous system than in the homogeneous system.

In general, our results for spatially heterogeneous systems agree qualitatively with those for homogeneous systems, suggesting that the insights gained from the study of homogeneous systems have a broad range of validity.

\section*{Discussion}
%The Discussion should be succinct and must not contain subheadings.
Our study has shown that global top predators can have a considerable stabilizing effect on meta-foodwebs. This result was obtained for a broad range of models and parameter values, as we used a generalized modelling approach that does not depend on the specific functional form of the interaction terms. The average stability of meta-foodwebs with a global top predator was larger than that of systems where the top predator was made local. The difference is often of the order of a factor of two. This means that when a global predator is made local in a stable ecosystem, the chance is around 50 percent that the system becomes unstable. 
Furthermore, we found that the stability of meta-foodwebs increases when the global predator has more prey species, and in many cases exceeds that of a system that has no such top predator. This latter effect appears to be considerably stronger when the system is heterogeneous, i.e., when different habitats have different equilibrium biomass densities.

Ripple et al.\cite{Ripple2014} mention a variety of functions that large predators fulfill in ecosystems around the world. Top predators are important at controlling the abundance in the lower tropic levels, this includes mesopredators\cite{Ritchie2009} as well as herbivores. A decline in big top predators often results directly in a release of mesopredators. Indirect effects can cascade through the whole foodweb. Due to the stronger mesopredator populations the abundance of prey species on lower tropic levels may decrease. Despite the release of intermediate predators a lower number of top predators can lead to a limited control of herbivores and therefore to a decrease in plant abundance. Large top predators are also thought to play an important role for the resilience of ecosystems against invading species. Therefore a loss of the big top predators can have severe consequences especially in the context of climate change.

The extent to which habitats are fragmented has important effects on the functioning and stability of ecosystems \cite{Fischer2007}, and especially on top predators\cite{Ritchie2009}. Far-ranging predators are negatively affected by landscape fragmentation because they are exposed most strongly to edge effects, for instance due to conflicts with humans \cite{Woodroffe1998}. Our results show that the stabilizing effect of large top predators depends on their capability to range freely. But this is hindered by landscape fragmentation. Even if a top predator was able to survive the fragmentation of its habitat it could lose its stabilizing capabilities. Previous results also show how habitat fragmentation can completely change the trophic role of top predators\cite{Layman2007}.

Taken together, the results highlight the importance of far-ranging top predators for ecosystem stability. The decline of those predators in natural systems is a serious concern. The extinction of predatory mammals like the Eurasian brown bear, lynx and wolf on the British isle are only one of many examples. While some of these extinctions are caused by climatic and environmental changes, anthropogenic impacts are the biggest factor. The ecological effects of such island extinctions cascade across ecological networks and take centuries to be fully realized \cite{wood2017}. The protection or reintroduction of large top predators is therefore of great importance. However, our study suggests that their large home ranges must also be restored if they shall have a considerable stabilizing effect.

\section*{Methods}

\subsection*{Generalized Modeling}
 We start with the general model equations (\ref{eqn:biomass_ode1}) and (\ref{eqn:biomass_ode2}), assuming that they have a steady state. We normalize all biomass densities and functions to this steady state. For example, if the population density of a local species at the steady state is denoted as $X_i^{k*}$ then the normalized population density is $x_i^k = \frac{X_i^k}{X_i^{k*}}$, and the normalized growth function is $g(x_i^k)=\frac{G(X_i^k)}{G(X_i^{k*})}$. This process yields the normalized meta-foodweb model equations 
\begin{align}
	\dot x_i^k = \alpha_{X_i^k}
	\Bigg[
		&\quad \, \tilde \nu_{X_i^k} \tilde\delta_{X_i^k} g_{X_i^k}(x_i^k) 
		+ \tilde\nu_{X_i^k} \delta_{X_i^k} f_{X_i^k} (t_{X_i^k},x_i^k) 
		- \tilde \rho_{X_i^k} \tilde\sigma_{X_i^k} m_{X_i^k}(x_i^k) 
		- \tilde\rho_{X_i^k} \sigma_{X_i^k} \tilde\xi_{X_i^k} \sum_j \beta_{X_j^k X_i^k} d_{X_j^k X_i^k}(\mathbf x^k) \notag\\
		&+ \sum_l \nu_{X_i^k X_i^l} e_i^{kl}(\mathbf x^k,\mathbf x^l) 
		- \sum_l \rho_{X_i^l X_i^k} e_i^{lk}(\mathbf x^l, \mathbf x^k)  
		- \tilde\rho_{X_i^k} \sigma_{X_i^k} \xi_{X_i^k} \sum_j \beta_{Y_j X_i^k} d_{Y_j X_i^k}(\mathbf x, \mathbf y)
	\Bigg] \,, \label{eqn:normalized_model_x}
\end{align}
where $i=1,\ldots,S$ and $k=1,\ldots,N$, and
\begin{align}
	\dot y_i = \alpha_{Y_i}
	\Bigg[\tilde\delta_{Y_i} g_{Y_i}(y_i) + \delta_{Y_i} f_{Y_i}(t_{Y_i},y_i) - \tilde \sigma_{Y_i} m_{Y_i}(y_i) - \sigma_{Y_i} \sum_j \beta_{Y_j Y_i} d_{Y_j Y_i}(\mathbf x, \mathbf y) \Bigg] \,, \label{eqn:normalized_model_y}
\end{align}
with $i=1,\ldots,S'$. The factors in front of the normalized functions are expressed in terms of scale parameters, the interpretation of which are listed in Table~\ref{tab:GeneralizedParameters}. Apart from the $\alpha$, which are the biomass flow rates, the other scale parameters are dimesionless and lie in the interval $[0,1]$.  In our model, the scale parameters are related to the structure of the spatial topology and of the local foodweb, as detailed in the next section.

The first step in a linear stability analysis consists in calculating  the Jacobian. In addition to the scale parameters, this Jacobian depends also on exponent parameters, which are the derivatives of the normalized functions with respect to  the normalized biomass densities \cite{Gross2006}. The exponent parameters (or elasticities) for the dynamics within a patch are 
\begin{align}
	\phi_{X_i^k} &= \left. \frac{\partial}{\partial x_i^k} g_{X_i^k}(x_i^k) \right|_{x=x^*} \,, &
    \phi_{Y_i} &= \left. \frac{\partial}{\partial y_i} g_{Y_i}(y_i) \right|_{y=y^*} \,, \nonumber\\
	\mu_{X_i^k} &= \left. \frac{\partial}{\partial x_i^k} m_{X_i^k}(x_i^k) \right|_{x=x^*} \,, &
    \mu_{Y_i} &= \left. \frac{\partial}{\partial y_i} m_{Y_i}(y_i) \right|_{y=y^*} \,, \nonumber\\
	\lambda_{X_j^k X_i^k} &= \left. \frac{\partial}{\partial x_i^k} r_{X_j^k X_i^k}(x_i^k) \right|_{x=x^*} \,, &
    \lambda_{Y_j Y_i} &= \left. \frac{\partial}{\partial y_i} r_{Y_j Y_i}(y_i) \right|_{y=y^*} \,, \nonumber\\
	\lambda_{Y_j X_i^k} &= \left. \frac{\partial}{\partial x_i^k} r_{Y_j X_i^k}(x_i^k) \right|_{x=x^*} \,, \nonumber\\
	\gamma_{X_i^k} &= \left. \frac{\partial}{\partial t_{X_i^k}} f_{X_i^k}(t_{X_i^k},x_i^k) \right|_{x=x^*} \,, &
    \gamma_{Y_i} &= \left. \frac{\partial}{\partial t_{Y_i}} f_{Y_i}(t_{Y_i},y_i) \right|_{y=y^*} \,, \nonumber\\
	\psi_{X_i^k} &= \left. \frac{\partial}{\partial x_i^k} f_{X_i^k}(t_{X_i^k},x_i^k) \right|_{x=x^*} \,, &
    \psi_{Y_i} &= \left. \frac{\partial}{\partial y_i} f_{Y_i}(t_{Y_i},y_i) \right|_{y=y^*} \,.
\end{align}
The exponent parameters related to dispersal between patches are
\begin{align}
	\hat \omega_{X_i^k X_i^l} &= \left. \frac{\partial}{\partial x_i^k} e_i^{kl}(x^k_1,\ldots,x^k_S,x^l_1,\ldots,x_S^l) \right|_{x=x^*} \,, &
	\omega_{X_i^k X_i^l} &= \left. \frac{\partial}{\partial x_i^l} e_i^{kl}(x^k_1,\ldots,x^k_S,x^l_1,\ldots,x_S^l) \right|_{x=x^*} \,, \nonumber\\
	\hat \kappa_{X_i^k X_j^l} &= \left. \frac{\partial}{\partial x_j^k} e_i^{kl}(x^k_1,\ldots,x^k_S,x^l_1,\ldots,x_S^l) \right|_{x=x^*} \,, & 
	\kappa_{X_i^k X_j^k} &= \left. \frac{\partial}{\partial x_j^l} e_i^{kl}(x^k_1,\ldots,x^k_S,x^l_1,\ldots,x_S^l) \right|_{x=x^*} \,,
\end{align}
where $i\neq j$. The interpretation of the exponent parameters can be found as well in Table~\ref{tab:GeneralizedParameters}. Note that different steady states are characterized by different values of the scale and exponent parameters and thus can have different stability properties. 

Although the functions and the steady states are unknown, the exponent  parameters have an clear meaning in the context of the system, and their typical range of values can therefore be obtained from general considerations (see Gross \textit{et al.}\cite{Gross2006,Gross2009}). 
This range of values is given in Table~\ref{tab:ModelParameters} for three types of models. The first model is shaped after systems with a Holling type 2 functional response and conventional, diffusive dispersal. The Standard model uses the largest meaningful range of exponent parameters, as specified in previous publications\cite{Gross2009,Plitzko2012,Gramlich2015,Gramlich2018}. The last model implements an adaptive type of dispersal, where the prey avoids its predators and the predators follow their prey.

\subsection*{Generation of meta-foodwebs}

The spatial topologies used in this paper were random geometric graphs (RGGs)\cite{Gilbert1961}. These graphs were generated by randomly placing $N$ patches on a unit square and connecting all pairs of patches that had a distance smaller than 
\begin{align}
	r=\frac{r'}{\sqrt N} \,,
\end{align}
with the parameter $r'$ radomly chosen between 1.4 and 2 for each RGG. This was done to cover an ensemble with a wider range of mean degrees. For the $N=10$ patch RGGs used in most of the simulations mean degrees range from approximately 1.8 to 9. The RGGs were generated with periodic boundary conditions to avoid edge effects. Only connected graphs were retained.

In most of the simulations we used spatial webs with $N=10$ patches. Test runs with larger numbers of patches showed no qualitative change in the results. A larger number of patches leads to a denser spectrum of the Laplacian eigenvalues, while the range of the eigenvalues is comparable.

If not specified otherwise we used the niche model \cite{Niche} to create the feeding relations between the different species. This model assigns to each species a random niche value $n_i$ between 0 and 1 using the uniform distribution.  We used a connectance of 0.25 and retained only connected niche webs. For systems  with a global species, the species with the largest niche value was identified as the global predator.
Niche values are correlated with bodymass, and we therefore implemented allometric scaling of the biomass turnover rate
\begin{align}
	{\alpha}_{i} &\sim 10^{-2n_i} \,.
\end{align}

The structures of the spatial habitat network and the local foodweb imposes relations between several scale parameters: 
The total biomass turnover rate consists of the turnover due to dispersal between the spatial patches and the turnover due to local in patch dynamics including the interaction with the global predator. To obtain homogeneous steady states we assume linkwise dispersal. This means all the links between patches have the same biomass turnover rate. Therefore the biomass turnover due to dispersal in a patch is proportional to the number of links. As the number of incoming and outgoing links is identical for undirected spatial graphs, the biomass gain and loss by dispersal cancel each other out ($\nu = \rho$). We fulfil all these conditions by using 
for each species a random factor $a_i \in [0,1]$ to fix the local biomass turnover
\begin{align*}
    \tilde \rho_i \alpha_i = \tilde \nu_i \alpha_i = a_i \cdot 10^{-2n_i}
\end{align*}
and the biomass turnover through one spatial link
\begin{align*}
    \rho_i^{kl} \alpha_i = \nu_i^{kl} \alpha_i = (1-a_i) \cdot 10^{-2n_i} \,.
\end{align*}
The remaining scale parameters are obtained via the following considerations:
In our model, the basal species gain biomass exclusively due to primary production ($\delta=0$). The only biomass input of the top and intermediate species is provided by feeding on their respective prey species ($\delta=1$). In order to fix the values of the scale parameters $\beta$ and $\chi$ in a consistent manner, we assign first a random link strength $s_{ij}$ to each feeding link between predator $i$ and prey $j$ using the standard normal distribution. This also includes the feeding links of the global predator. We interpret this link strength as the biomass flowing through link $ij$ per unit time. The scale parameters $\beta$ and $\chi$ give the relative contribution of the different feeding links to the biomass loss resp.~gain to a node of the foodweb, and they are therefore obtained from the link strengths $s_{ij}$ via the relations 
\begin{align} \label{eqn:feeding_scale}
	\beta_{ij} &= \frac{s_{ij}}{\sum_n s_{nj}} \,, &
	\chi_{ij} &= \frac{s_{ij}}{\sum_n s_{in}} \,,
\end{align}
where for local species the sums are calculated over only the local species. A sample foodweb with 5 species is shown in figure~\ref{fig:local5}. In order to distinguish between local and global losses the scale parameter $\xi$ is introduced. The fraction of loss by predation due to the global predator (index $g$) is calculated by
\begin{align*}
    \xi_i = \frac{s_{g i}}{\sum_n s_{n i}} \,,
\end{align*}
with the sum over all species including the global predator. The fraction of biomass loss due to predation $\sigma$ is randomly chosen between 0 and 1 for all prey species separately, for top species $\sigma=0$. 

Finally, the non-constant exponent parameters are drawn randomly from uniform distributions in the intervals given in table~\ref{tab:ModelParameters}, using the mentioned three models.

\subsection*{Structure of the Jacobian}

The Jacobian of the model represented by the equations (\ref{eqn:biomass_ode1}) and (\ref{eqn:biomass_ode2}) or, equivalently, by the equations (\ref{eqn:normalized_model_x}) and (\ref{eqn:normalized_model_y}),
has the structure (\ref{eqn:general_jacobian}),
which is illustrated in terms of smaller blocks of sizes $S\times S$, $S\times S'$, and $S' \times S'$ in figure \ref{fig:jacobian}.
The first block $\mathbf J_X$ contains the information about the dynamics of the $S$ local species and the matrix $\mathbf J_Y$ captures the dynamics of the $S'$ global species, while the matrices $\mathbf J_{XY}$ and $\mathbf J_{YX}$ capture the interaction between the spatial web and the global species due to global predation. 

The first block $\mathbf J_X$ itself can be constructed out of $N\times N$ block matrices of size $S\times S$. We denote their constituents as ${\mathbf P_X}^k$, which capture the local dynamics within patch $k$,  ${\mathbf C_X}^{kl}$ and ${\mathbf{\hat C}_X}{}^{kl}$, which capture the contributions to the Jacobian due to dispersal from patch $l$ to $k$, and  ${\mathbf O_X}^{kk}$ and ${\mathbf O_X}^{kl}$, which are nonzero when there is an indirect coupling of species within patch $k$ or between patches $k$ and $l$ due to nonlinear predation by global predators. With this notation, the first block of the Jacobian takes the form

\begin{align}
	\mathbf J_X = \begin{pmatrix}
		\ddots & \vdots & & \vdots\\
		\cdots & {\mathbf O_X}^{ik} & \cdots & {\mathbf O_X}^{il} & \cdots \\
		& \vdots & & \vdots \\
		\cdots & {\mathbf P_X}^k - \sum_m {\mathbf C_X}^{km} + {\mathbf O_X}^{kk} & \cdots & + {\mathbf{\hat C}_X}{}^{kl} + {\mathbf O_X}^{kl} & \cdots \\
		& \vdots & \ddots & \vdots \\
		\cdots &  {\mathbf{\hat C}_X}{}^{lk} + {\mathbf O_X}^{lk} & \cdots & {\mathbf P_X}^l - \sum_m {\mathbf C_X}^{lm} + {\mathbf O_X}^{ll} & \cdots \\
		& \vdots & & \vdots \\
		\cdots & {\mathbf O_X}^{jk} & \cdots & {\mathbf O_X}^{jl} & \cdots \\
		& \vdots & & \vdots & \ddots
	\end{pmatrix} \,.
\end{align}

The structure of the spatial network web enters via the positions of the coupling matrices $\mathbf C_X^{kl}$ and $\mathbf{\hat C}_X^{kl}$. In the case of a spatially homogeneous steady state all generalized parameters are are equal between the different patches. Therefore the different matrices occuring in the Jacobian do in fact not depend on the patch indices, and the matrices $\mathbf C_X$ and $\mathbf{\hat C}_X$ become equal. Using the identity matrix $\mathbf{I}$, the Laplacian matrix of the spatial patch network $\mathbf L$, and the matrix $\mathbf N$, that is 0 on the diagonal and 1 everywhere else, the first block of the Jacobian can be written more compactly as
\begin{align}
	\mathbf J_X = \mathbf I \otimes \mathbf P'_X - \mathbf L \otimes \mathbf C_X + \mathbf N \otimes \mathbf O_X \,,
    \label{eqn:Jx_homogeneous}
\end{align}
where $\mathbf P'_X = \mathbf P_X + {\mathbf O_X}^{kk}$. The remaining entries of  the matrix in figure \ref{fig:jacobian} are  $\mathbf J_Y$, which captures the dynamics of the global species, and the matrices ${\mathbf C_{XY}}^k$ and ${\mathbf C_{YX}}^k$, which capture the coupling of the spatial patches to the global species due to predation of local species by global species. 

\subsection*{Calculation of the eigenvalues}
The eigenvalues are calculated in a similar way as in \cite{Brechtel2018}.
The system has two types of eigenvectors, the homogeneous and the inhomogeneous ones. The homogeneous eigenvectors have the same components for every spatial patch and a non-zero vector component corresponding to the global predator. The components of the inhomogeneous eigenvectors differ between the spatial patches while the vector component corresponding to the global predator vanishes. In the following, we calculate these two types of eigenvectors separately.
\par
We start with the inhomogeneous eigenvectors by calculating the eigenvalues of the first block of the Jacobian $\mathbf J_X$. In the next step we show that certain eigenvalues of $\mathbf J_X$ are also eigenvalues of the full Jacobian $\mathbf J$.
\par
The eigenvalues of $\mathbf J_X$ have to be calculated from equation~(\ref{eqn:Jx_homogeneous}). The main difference to the system without global predators is given by the coupling matrices $\mathbf O_X^{kl}$ and $\mathbf O_X^{kk}$ and is caused by the indirect coupling between local species due to global predation. The Laplacian $\mathbf L$ and the matrix $\mathbf N$ can be written as
\begin{align}
	L_{kj} &= g_k \delta_{kj} - A_{kj} \,, &
    N_{ik} &= 1 - \delta_{ik} \,,
\end{align}
with $g_i$ being the degree of the vertex with index $i$,  $A_{kj}$ being the adjacency matrix, and $\delta_{ij}$ being the Kronecker delta. Both matrices commute, so there is a common basis of eigenvectors,
\begin{align}
	\mathbf L &= \kappa \mathbf p \,, &
	\mathbf N &= \mu \mathbf p \,.
\end{align}\\
This allows us to replace $\mathbf L$ and  $\mathbf N$ in (\ref{eqn:Jx_homogeneous}) by their eigenvalues, giving the following reduced eigenvalue equation
\begin{align}
	\label{eqn:reduced_eigenvalues}
	({\mathbf P}'_X - \kappa {\mathbf C}_X + \mu {\mathbf O}_X) \cdot \mathbf{q} = \lambda_{\kappa\mu} \mathbf{q}\, .
\end{align}
The full eigenvector can than be written as
\begin{align}
	v = \mathbf p \otimes \mathbf q \,,
\end{align}
and the complete eigenvalue equation of the first block of the Jacobian is then
\begin{align}
	\mathbf J_X \cdot \mathbf{v} 
	&= (\mathbf I \otimes \mathbf P'_X - \mathbf L \otimes \mathbf C_X + \mathbf N \otimes \mathbf O_X) \cdot (\mathbf{p} \otimes \mathbf{q}) \nonumber\\
	&= \mathbf I \cdot \mathbf{p} \otimes \mathbf P'_X \cdot \mathbf{q} - \mathbf L \cdot \mathbf{p} \otimes \mathbf C_X \cdot \mathbf{q} + \mathbf N \cdot \mathbf{p} \otimes \mathbf O_X \cdot \mathbf{q} \nonumber\\
	&= \mathbf{p} \otimes \mathbf P'_X \cdot \mathbf{q} - \kappa \mathbf{v} \otimes \mathbf C_X \cdot \mathbf{q} + \mu \mathbf{p} \otimes \mathbf O_X \cdot \mathbf{q} \nonumber\\
	&= \mathbf{p} \otimes (\mathbf P'_X - \kappa \mathbf C_X + \mu \mathbf O_X) \cdot \mathbf{q} \nonumber\\
	&= \mathbf{p} \otimes \lambda_{\kappa\mu} \mathbf{q} = \lambda_{\kappa\mu} (\mathbf{p} \otimes \mathbf{q}) =\lambda_{\kappa\mu} \mathbf{v}\,. \label{proof}
\end{align} 

Next we show that certain eigenvalues of $\mathbf J_X$ are also eigenvalues of $\mathbf J$. With a given eigenvector $\mathbf v$ corresponding to the eigenvalue $\lambda_{\kappa\mu}$ of $\mathbf J_X$ we find
\begin{align}
	\begin{pmatrix}
		\mathbf J_X & \mathbf J_{XY} \\
		\mathbf J_{YX} & \mathbf J_Y
	\end{pmatrix}
	\cdot
	\begin{pmatrix}
		\mathbf v \\
		0
	\end{pmatrix}
	&=
	\begin{pmatrix}
		\mathbf J_X \cdot \mathbf v \\
		\mathbf J_{YX} \cdot \mathbf v
	\end{pmatrix}
	=
	\begin{pmatrix}
		\lambda_{\kappa\mu} \mathbf v \\
		\mathbf J_{YX} \cdot \mathbf v
	\end{pmatrix} \, .
\end{align}
The constructed vector $(\mathbf v, 0)^T$ is an eigenvector if 
\begin{align}
	\mathbf J_{YX} \cdot \mathbf v \overset != 0 \,.
\end{align}
The Jacobians block in question $\mathbf J_{YX}$ has the structure
\begin{align}
	\mathbf J_{YX} = (1,\dots,1)\otimes \mathbf C_{YX} \,.
\end{align}
Therefore it follows
\begin{align}
	\mathbf J_{YX} \cdot \mathbf v
	&= [(1,\dots,1)\otimes \mathbf C_{YX}] \cdot (\mathbf p \otimes \mathbf q) \notag \\
	&= \left[ (1,\dots,1) \cdot \mathbf p \otimes \mathbf C_{YX}\cdot \mathbf q \right] \,.
\end{align}
Accordingly we only need to show that
\begin{align}
	\label{eqn:vector_sum_zero}
	(1,\dots,1) \cdot \mathbf p \overset != 0 \,.
\end{align}
We multiply the eigenvector equation from the left side with a row vector filled with 1,
\begin{align}
	(1,\dots,1) \cdot \mathbf L &= \kappa \, (1,\dots,1) \cdot \mathbf{p} \,.
\end{align}
Because $(1,\dots,1) \cdot \mathbf L = 0$, either $\kappa = 0$ or $(1,\dots,1) \cdot \mathbf{p} = 0$. 
Consequently the condition holds for the eigenvectors with $\kappa \neq 0$.
\par
The multiplicity of the eigenvalue $\kappa=0$ of the Laplacian is the number of connected components of the considered spatial graph. In this paper we focus only on connected graphs, therefore the multiplicity of $\kappa=0$ is 1. The matrix $\mathbf N$ has the eigenvalue $\mu = N-1$ exactly once, and the only other eigenvalue is $\mu = -1$. With this eigenvalue, we have
\begin{align}
	\mathbf N \cdot \mathbf p &= - \mathbf p\, .
\end{align}
We multiply this equation again from the left side with a row vector filled with 1 and rearrange the equation,
\begin{align}
	(1,\dots,1) \cdot \mathbf N \cdot \mathbf p &= -(1,\dots,1) \cdot \mathbf p \nonumber\\
	(N-1,\dots,N-1) \cdot \mathbf p &= -(1,\dots,1) \cdot \mathbf p \nonumber\\
	(N-1) \, (1,\dots,1) \cdot \mathbf p &= -(1,\dots,1) \cdot \mathbf p \nonumber\\
	N \, (1,\dots,1) \cdot \mathbf p &= 0 \nonumber\\
	(1,\dots,1) \cdot \mathbf p &= 0\, .
\end{align}
Therefore the condition given by equation~\eqref{eqn:vector_sum_zero} holds also for eigenvectors with $\mu = -1$.
\par
The eigenvalues of $\mathbf J_X$ with $\kappa\neq0$ and the eigenvalues with $\mu=-1$, which we have calculated using equation~\eqref{eqn:reduced_eigenvalues} are also eigenvalues of $\mathbf J$. These eigenvalues and eigenvectors correspond to the inhomogeneous eigenmodes of the spatial system.
\par
Next we look at the spatially homogeneous eigenvectors. The Laplacian $\mathbf L$ and the matrix $\mathbf N$ share the common eigenvector $\mathbf p = (1,...,1)^T$ with the eigenvalue $\kappa = 0$ and $\mu = N-1$ respectively. This eigenvector corresponds to a homogeneous behavior of the spatial system which is influenced by the global species. Therefore we are unable to calculate the eigenvalue of the Jacobian corresponding to the homogeneous eigenmode by taking only $\mathbf J_X$ into account. We calculate  the corresponding eigenvalues by simplifying the eigenvalue equation of the Jacobian to the homogeneous case by using eigenvectors constructed as
\begin{align}
	\mathbf{v} = 
	\begin{pmatrix}
		\mathbf{1} \otimes \mathbf{v}_X \\
		\mathbf{v}_Y
	\end{pmatrix} \, ,
\end{align}
where
\begin{align}
        \mathbf{1} =
        \begin{pmatrix}
			1\\
			\vdots\\
			1
		\end{pmatrix}
\end{align}
and $\text{dim}(\mathbf 1) = N$. In order to show that $\mathbf{v}$ is indeed an eigenvector, we multiply it with the complete Jacobian:
\begin{align}
	&
	\begin{pmatrix}
		\mathbf I \otimes \mathbf P'_X - \mathbf L \otimes \mathbf C_X + \mathbf N \otimes \mathbf O_X &
		\mathbf{1} \otimes \mathbf{C}_{XY} \\
		(1,\dots,1) \otimes \mathbf{C}_{YX} &
		\mathbf{P}_Y
	\end{pmatrix}
	\begin{pmatrix}
		\mathbf{1} \otimes \mathbf{v}_X \\
		\mathbf{v}_Y
	\end{pmatrix}
	=
	\lambda 
	\begin{pmatrix}
		\mathbf{1} \otimes \mathbf{v}_X \\
		\mathbf{v}_Y
	\end{pmatrix} \nonumber\\
	=&
	\begin{pmatrix}
		\mathbf I \cdot \mathbf{1} \otimes \mathbf P'_X \cdot \mathbf{v}_X - \mathbf L \cdot \mathbf{1} \otimes \mathbf C_X \cdot \mathbf{v}_X + \mathbf N \cdot \mathbf{1} \otimes \mathbf O_X \cdot \mathbf{v}_X + \mathbf{1}
		\otimes \mathbf{C}_{XY} \cdot \mathbf{v}_Y \\
		(1,\dots,1) \cdot \mathbf{1} \otimes \mathbf{C}_{YX} \cdot \mathbf{v}_{X} + \mathbf{P}_{Y} \cdot \mathbf{v}_{Y}
	\end{pmatrix} \nonumber\\
	=&
	\begin{pmatrix}
        \mathbf{1}
		\otimes
		\left[
			\mathbf P'_X \cdot \mathbf{v}_X
			+ (N-1) \mathbf O_X \cdot \mathbf{v}_X
			+ \mathbf{C}_{XY} \cdot \mathbf{v}_Y 
		\right]\\
		N \mathbf{C}_{YX} \cdot \mathbf{v}_X + \mathbf{P}_Y \cdot \mathbf{v}_Y
	\end{pmatrix}
	=
	\lambda 
	\begin{pmatrix}
		\mathbf{1} \otimes \mathbf{v}_X \\
		\mathbf{v}_Y
	\end{pmatrix} \,.
\end{align}

In summary, we reduced the eigenvector equation of the spatially homogeneous eigenmodes to
\begin{align}
	\begin{pmatrix}
		\mathbf P'_X + (N-1) \mathbf O_X &
		\mathbf{C}_{XY}\\
		N \mathbf{C}_{YX} & \mathbf{P}_Y
	\end{pmatrix}
	\begin{pmatrix}
		\mathbf{v}_{X} \\
		\mathbf{v}_{Y}
	\end{pmatrix}
	=
	\lambda 
	\begin{pmatrix}
		\mathbf{v}_{X} \\
		\mathbf{v}_{Y}
	\end{pmatrix} \,.
\end{align}
The dimension of this eigenvalue equation does not depend on the size of the spatial web. The spatial web enters only via the number of patches $N$. In addition the equation does not depend on the spatial coupling $\mathbf C_X$ between local species due to dispersal. 
\par
This eigenvalue equation gives us the spatially homogeneous eigenmodes, where all the patches behave the same way. The local populations do not exchange biomass between patches. The global predators only react to the total available biomass in the complete system and are therefore insensitive to the topology of the spatial web. Hence the coupling between global and local species is described by the local eigenmode of the system.

\section*{Ensemble sizes}
For the systems in figures \ref{fig:PSWbarplotHom}, \ref{fig:PSWbarplotHom2} and \ref{fig:stab-gen-vs-spec} we used ensembles of $10^6$ webs per class.

\begin{figure*}
	\centering
	\includegraphics[width=\textwidth]{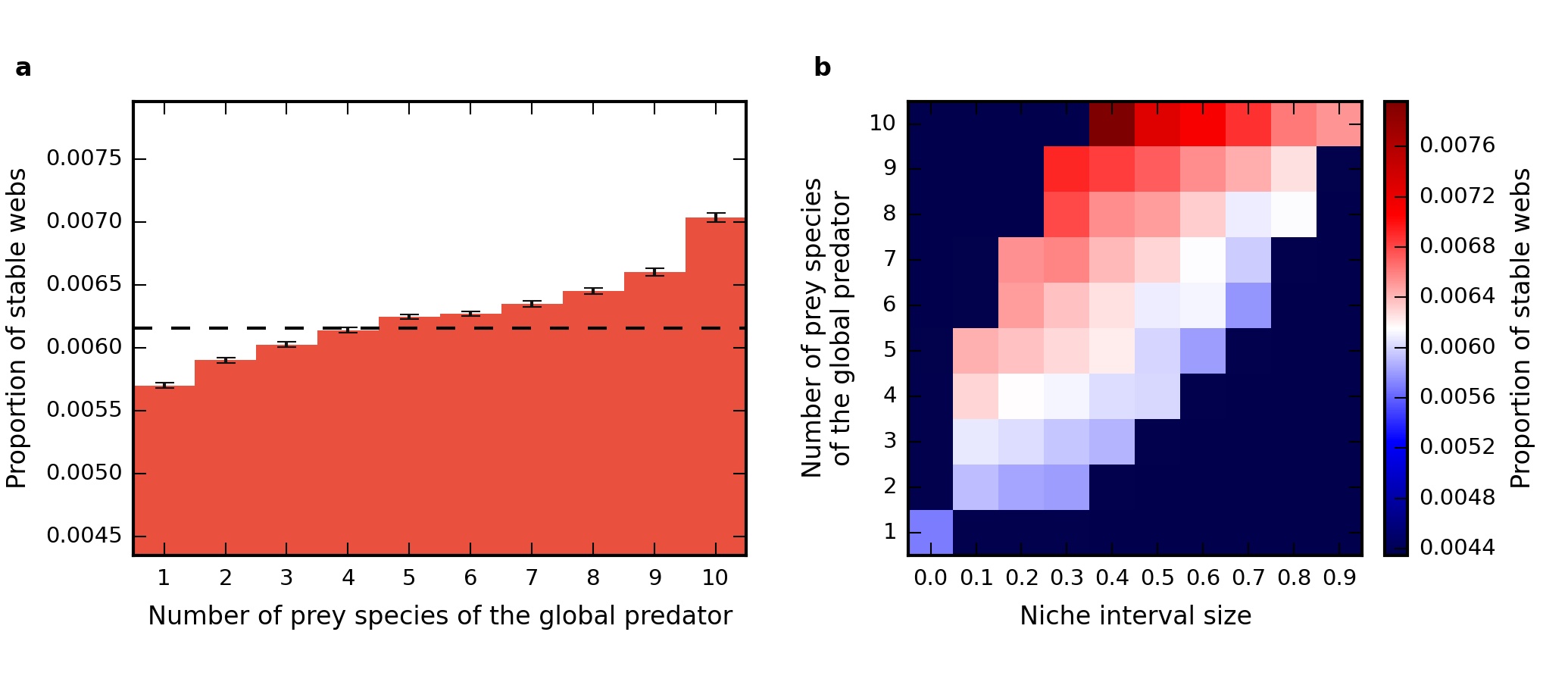}
	\caption{\textsf{\textbf{a.}} Proportion of stable webs for a 10-patch Holling type 2 system with $S=10$ local species and a global predator for the different numbers of prey species of the global predator. The dashed line represents the proportion of stable webs of the system when the global species is removed. The system's stability increases when the global species has more prey species. For more than 4 prey species the system with the global predator exceeds the stability of the system without it. \textsf{\textbf{b.}} Proportion of stable webs for the same system as in figure~\ref{fig:LinkNumberHist+NicheIntervalHist2D}~a, showing additionally he dependence on the niche interval size. The niche interval size is the difference between the largest and the smallest niche values of global predator's prey species. White color corresponds to the proportion of stable webs of local system when the global species is removed, red color corresponds to higher and blue color to lower stability.}
	\label{fig:LinkNumberHist+NicheIntervalHist2D}
\end{figure*}

\begin{figure*}
	\centering
    \includegraphics{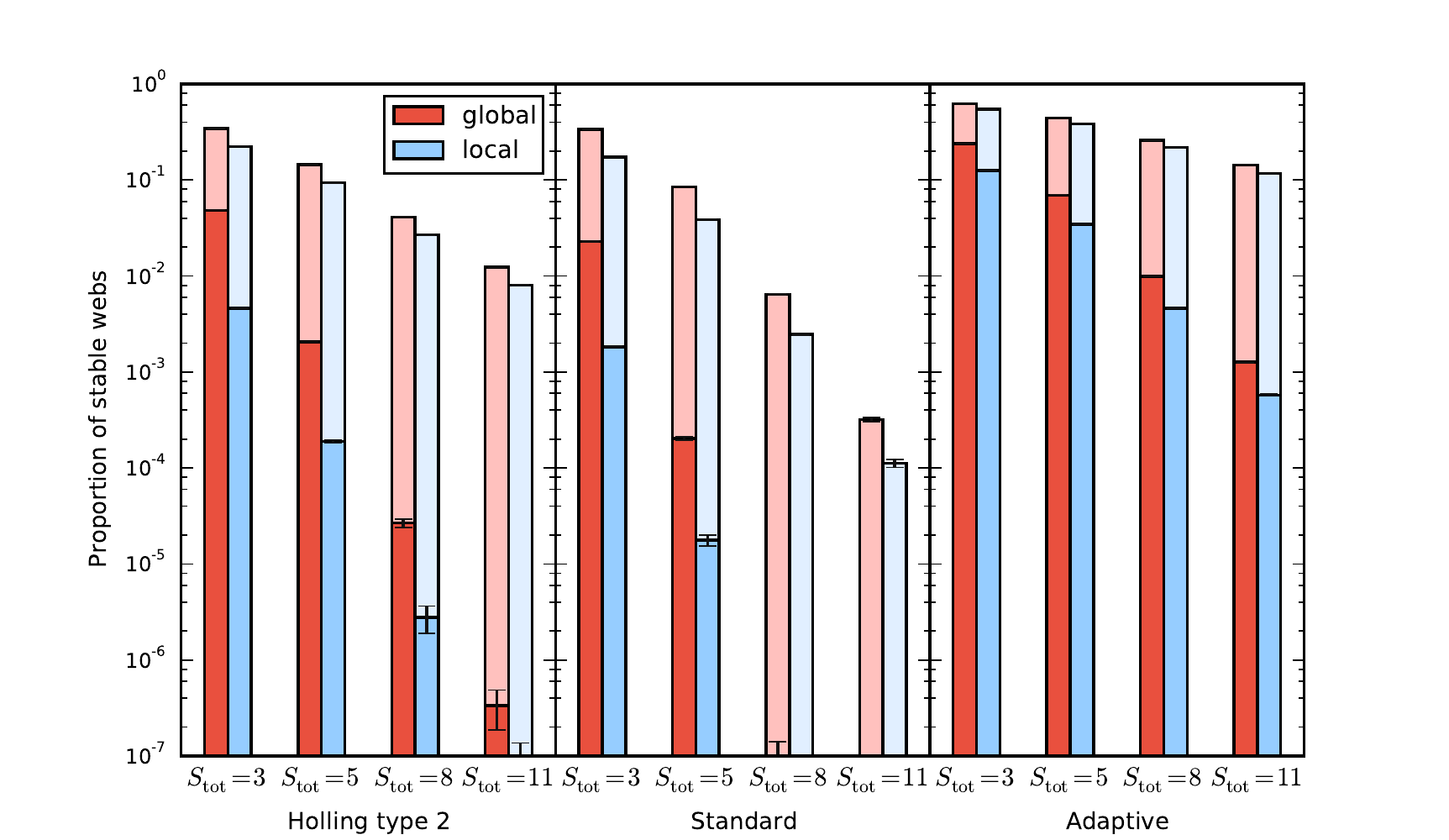}
	\caption{Proportion of stable webs for heterogeneous five-patch systems with a global top predator (red) and systems with a local top predator (blue), for the three models specified by the parameter intervals listed in Table \ref{tab:ModelParameters}, and for different species numbers $S_{\rm{tot}}=S+S'$. The figure shows that making a global predator local decreases stability also for the more general, inhomogeneous meta-foodwebs. The lighter shaded bars indicate the proportion of stable webs of the corresponding homogeneous systems. In the model shown here, the heterogenoeus meta-foodwebs have a lower average stability than their homogeneous counterparts. 
	}
	\label{fig:PSWbarplotHet}
\end{figure*}

\begin{figure*}
	\centering
    \includegraphics{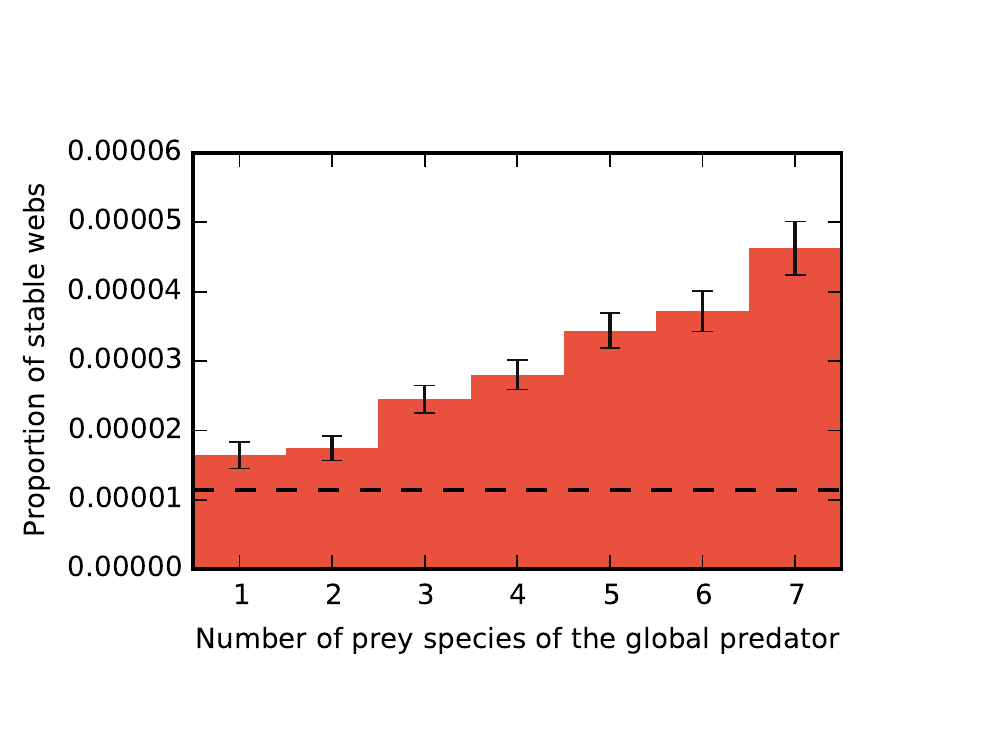}
	\caption{Proportion of stable webs for heterogeneous five-patch systems with a global top predator and $S=7$ local species for the Holling 2 specified by the parameter intervals listed in Table \ref{tab:ModelParameters}. The figure shows that a higher number of prey species leads to more stable systems. This is in agreement with the homogeneous case (see figure~\ref{fig:LinkNumberHist+NicheIntervalHist2D}~a). The stability of the system without the global top predator is indicated by the dashed line, showing that it is less stable than the system with global top predator.}
	\label{fig:LinkNumberHistHet}
\end{figure*}

\begin{figure}
	\begin{center}
		\includegraphics[width=0.4\textwidth]{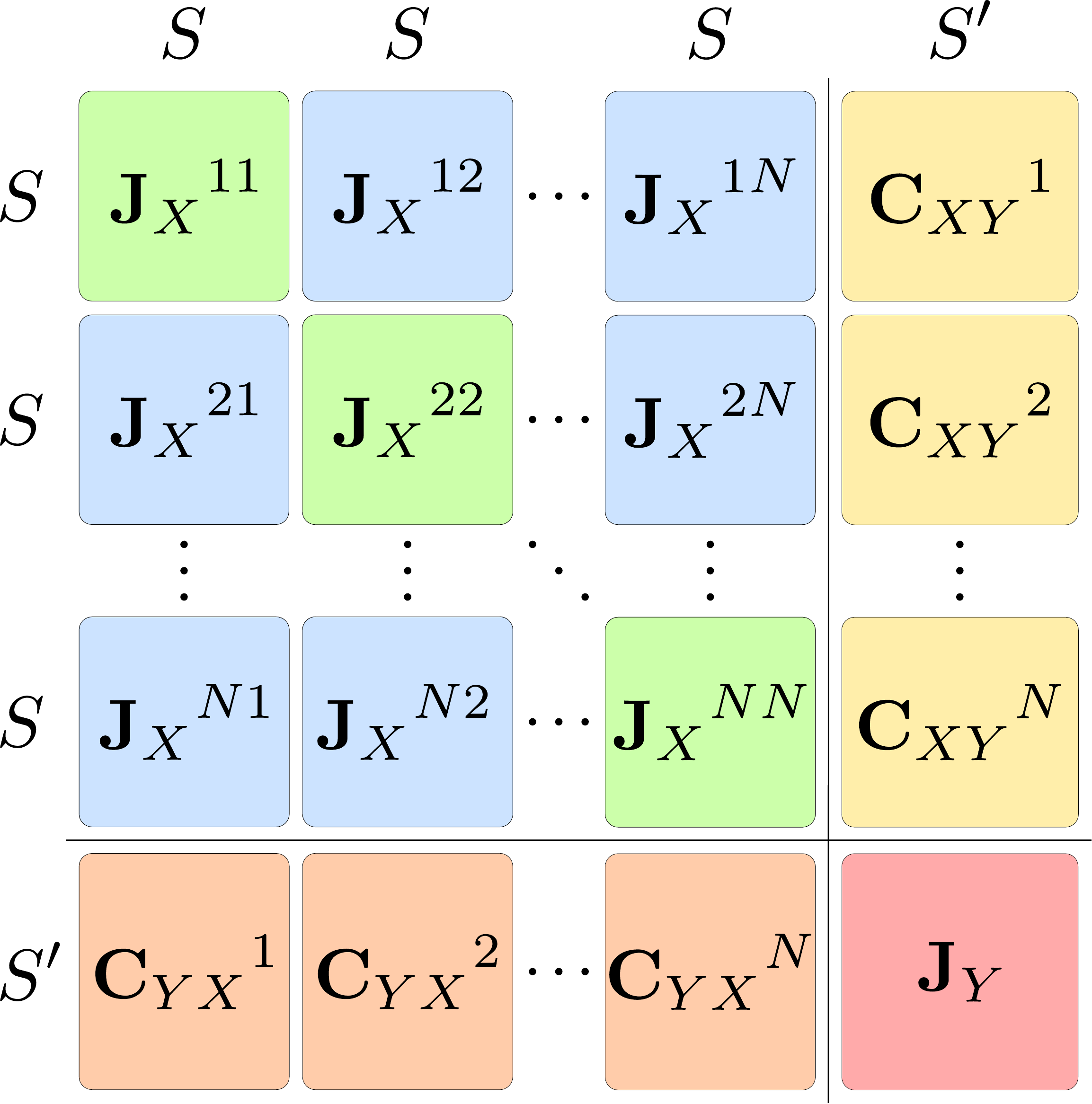} 
    \end{center}
    \caption{Structure of the Jacobian given in  equation (\ref{eqn:general_jacobian}). The block $\mathbf J_X$ consists of the blue and green parts, the block $\mathbf J_{XY}$ consists of the yellow parts, the block $\mathbf J_{YX}$ consists of the orange blocks and the block $\mathbf J_Y$ is shown in red.}
    \label{fig:jacobian}
\end{figure}

\begin{figure}
	\centering
	\includegraphics[scale=1]{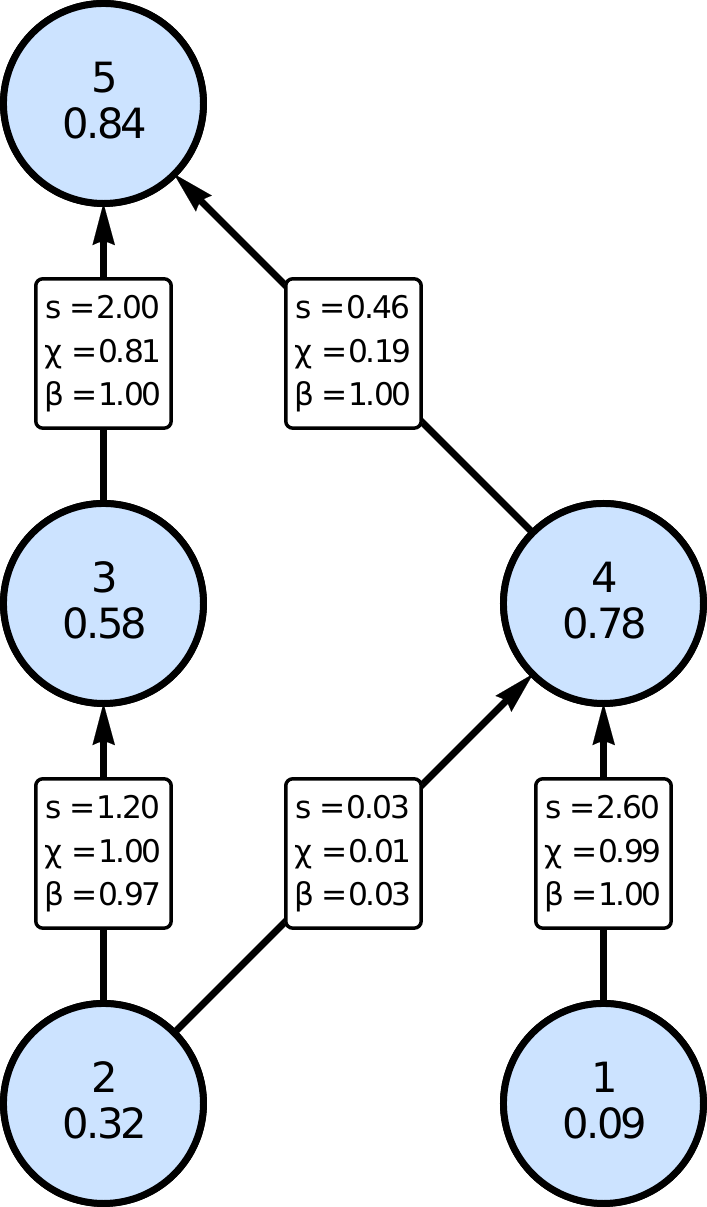}
	\caption{Illustration of the assignment of the scale parameters for a local foodweb with 5 species. In the circles, the species index and the niche value are given. On the links, the relative link strength $s$ is given as well as the scale parameters $\chi$ and $\gamma$. According to the relations (\ref{eqn:feeding_scale}) the parameters $\beta$ add up to 1 for a given prey (e.g., species number 2  being eaten by 3 and 4), and the parameters $\chi$ add up 1 for a given predator  (e.g., species 4 feeding on 1 and 2). }
	\label{fig:local5} 
\end{figure}

In figure~\ref{fig:LinkNumberHist+NicheIntervalHist2D} we used an ensemble of more than $10^7$ webs. The stability baseline of the system without the global predator in figure~\ref{fig:LinkNumberHist+NicheIntervalHist2D}~a was calculated with an ensemble of $10^6$ webs. In figure \ref{fig:LinkNumberHist+NicheIntervalHist2D}~b only bins with sufficient statistics are shown. Statistics gets worse for more extreme combinations of prey numbers and niche interval sizes.

For the systems in figure \ref{fig:PSWbarplotHet} we used ensembles of various sizes ranging from $3.1\cdot 10^6$ for the more stable systems to over $1.3\cdot 10^7$ for the least stable systems. Note that for the three least stable classes not even one single stable web was found.

In figure~\ref{fig:LinkNumberHistHet} an ensemble of more than $3.5\cdot10^7$ webs was used. The stability baseline of the system without the global predator was calculated with an ensemble of $5\cdot 10^6$ webs.

The proportion of stable webs (PSW) is calculated by
\begin{align}
    \textsf{PSW} = \frac{N_\text{stable}}{N} \;,
\end{align}
where $N$ is the total number of tested webs and $N_\text{stable}$ is the number of stable webs found in the given sample set. The standard deviation of the PSW is estimated by
\begin{align}
    \Delta \text{PSW} = \frac{\sqrt{N_\text{stable}}}{N} \;.
\end{align}
Error bars are only shown when the standard derivation is visible on the scale of the plot. In figure \ref{fig:stab-gen-vs-spec} the error bars would be smaller than the line width used in the plot.

\section*{Data Availability}
This study did not involve any underlying data. All the information needed to generate the data analyzed during this study is included in this published article.
 
\bibliography{lit}

\section*{Acknowledgements}
We acknowledge support by the German Research Foundation (DFG research unit 1748, contract number Dr300/13-2) and by the EPSRC (EP/N034384/1). 
For coverage of the publication costs, we acknowledge support by the German Research Foundation and the Open Access Publishing Fund of Technische Universität Darmstadt.

\section*{Author information}
\subsection*{Affiliations}
Institute of Condensed Matter Physics, Technische Universität Darmstadt,\\
Darmstadt, Germany\\
Andreas Brechtel \& Barbara Drossel
\par
UC Davis, Department of Computer Science, 1 Shields Av\\ Davis, Ca 95616, USA\\
Thilo Gross

\subsection*{Contributions}
All authors contributed to and reviewed the manuscript. A.B. performed the simulations and prepared the figures.

\subsection*{Competing Interests}
The authors declare that they have no competing interests.

%Table parameter interpretations
\begin{table*}
\begin{tabular}{l|p{0.75\textwidth}}
\hline
Parameter & Interpretation\\
\hline
\hline
Exponent& \\
\hline
$\phi$ & Sensitivity of primary production to own population density\\
$\gamma$ & Sensitivity of predation to total available prey density \\   	
$\lambda$ & Exponent of prey switching\\
$\psi$ & Sensitivity of predation to predator density\\
$\mu$ & Exponent of closure\\
\hline
$\omega$ & Sensitivity of dispersal to own the species population density in the output patch\\
$\tilde{\omega}$ & Sensitivity of dispersal to own species population density in the target patch\\
$\kappa$ & Sensitivity of dispersal to population density of some other species in the output patch\\
$\hat{\kappa}$ & Sensitivity of dispersal the population density of some other species in the target patch\\
\hline
\hline
Scale & \\
\hline
${\alpha}$ & Biomass flow\\
$\sigma$ & Fraction of biomass loss due to predation\\
$\tilde{\sigma}$ & Fraction of biomass loss due to respiration\\
$\xi$ & Fraction of loss by predation due to global species\\
$\tilde{\xi}$ & Fraction of loss by predation due to local species\\
$\beta$ & Relative contribution to biomass loss due to predation by a certain predator\\
$\delta$ & Fraction of local growth by predation\\  
$\tilde{\delta}$ & Fraction of local growth by primary production\\  
$\chi$ & Relative contribution of population as prey to a certain predator\\
\hline
$\nu$ & Fraction of total biomass gain due to dispersal\\
$\tilde{\nu}$ & Fraction of total biomass gain due to predation and primary production\\
$\rho$ & Fraction of total biomass loss due to dispersal\\
$\tilde{\rho}$ & Fraction of total biomass loss due to predation and respiration\\
\hline
\end{tabular}
\caption{Generalized parameters used to describe the meta-foodweb.}
\label{tab:GeneralizedParameters}
\end{table*}

%Table parameter intervalls
\begin{table*}
	\centering
	\begin{tabular}{l c c c c c c c c c c c c c c}
	\hline
	& $\phi_X$ / $\phi_Y$ & $\psi_X$ / $\psi_Y$ & $\mu_X$ / $\mu_Y$ & $\gamma_X$ / $\gamma_Y$ & $\lambda_X$ / $\lambda_{YX}$ & $\omega_X$ & $\hat\omega_X$ & $\kappa_X$ & $\hat\kappa_X$ \\
	\hline
	\hline
	Holling T2 & 1 & 1 & [1,2] & [0,1] & [1,2] & [0,1] & [0,1] & 0 & 0 \\
    Standard & [0,1] & [0,1] & [1,2] & [0,2] & [1,2] & [-2,2] & [-2,2] & [-2,2] & [-2,2] \\
   Adaptive & [0,1] & [0,1] & [1,2] & [0,2] & [1,2] & [0,1] & [0,1] & [0,1] & [0,1] \\
	\hline
	\end{tabular}
	\caption{Constant values and distribution intervals of the exponent parameters. In the case of the 'Adaptive' model the parameters $\kappa_X$ and $\hat\kappa_X$ depend on the feeding links of the local foodweb. If species $i$ is the prey of species $j$ then $\kappa_{X_i X_j}$ is randomly assigned a value inside the interval $[0,1]$ and is otherwise set to 0. This means the dispersal of prey is stimulated if predators are present. Additionally the parameter $\hat\kappa_{X_j X_i}$ is randomly assigned a value inside the interval [0,1] if species $j$ does feed on species $i$ and is set to 0 otherwise. This corresponds to predators that prefer to disperse into patches with higher prey abundance.}
	\label{tab:ModelParameters}
\end{table*}

%Figure PSW-Barplot homogeneous

%Figure stability curves

%Figure LinkNumberHist + NicheIntervalHist2D

%Figure PSW-Barplot heterogeneous

%Figure LinkNumberHist heterogeneous

%Figure Jacobian structure

%Figure 5 species foodweb

\end{document}